\documentclass[a4paper,man,natbib,floatsintext]{apa6}
\usepackage{apacite}
\usepackage{subfig}
\usepackage[english]{babel}
\usepackage[utf8x]{inputenc}
\usepackage{amsmath}
\usepackage{graphicx}
\usepackage[colorinlistoftodos]{todonotes}
\usepackage[left]{lineno}
\usepackage{diagbox}
\usepackage{indentfirst}
\usepackage{adjustbox}
\usepackage{xcolor}
\usepackage{multirow}
\usepackage{caption}
\usepackage{setspace}
\begin{document}

\raggedbottom

%%To left align captions
\captionsetup[table]{
  labelsep = newline,
 % textfont = sc, 
 % name = TABLE, 
  justification=raggedleft,
  singlelinecheck=off,%%%%%%% a single line is centered by default
  labelsep=colon,%%%%%%
  skip = \medskipamount}

\begin{titlepage}

\begin{center}

\large Investigating Explanations in Conditional and Highly Automated Driving: The Effects of Situation Awareness and Modality
% Physiological perspective in takeover 
\\ 

\normalsize

\vspace{25pt}
Lilit Avetisyan\\
Industrial and Manufacturing Systems Engineering, University of Michigan-Dearborn\\
\vspace{15pt}
Jackie Ayoub\\
Industrial and Manufacturing Systems Engineering, University of Michigan-Dearborn\\
\vspace{15pt}
Feng Zhou\\
Industrial and Manufacturing Systems Engineering, University of Michigan-Dearborn\\
\vspace{15pt}
% Elizabeth M. Pulver\\
% State Farm Mutual Automobile Insurance Company\\
% \vspace{15pt}
% Dawn M. Tilbury\\
% Mechanical Engineering, University of Michigan\\
% \vspace{15pt}
% Lionel P. Robert\\
% School of Information, University of Michigan\\
% \vspace{15pt}
% Anuj K. Pradhan\\
% Industrial and Mechanical Engineering, University of Massachusetts Amhest\\
% \vspace{15pt}

\end{center}
\begin{flushleft}
%\textbf{\textit{Accepted to be published in Accident Analysis \& Prevention 09/23/2020}} \\
\textbf{Manuscript type:} \textit{Research Article}\\
\textbf{Running head:} \textit{The Effects of Situation Awareness and Modality}\\
\textbf{Word count:} %5800 \\ 

\textbf{Corresponding author:} 
Feng Zhou, 4901 Evergreen Road, Dearborn, MI 48128, Email: fezhou@umich.edu

%\textbf{Acknowledgment:} This work was supported by University of Michigan Mcity.
% The views expressed are those of the authors and do not reflect the official policy or position of State Farm\textsuperscript{\textregistered}.

\end{flushleft}

% \rule{\linewidth}{2pt}
% \vspace*{\stretch{2}}
\end{titlepage}
\shorttitle{}

\section{ABSTRACT}

With the level of automation increases in vehicles, such as conditional and highly automated vehicles (AVs), drivers are becoming increasingly out of the control loop, especially in unexpected driving scenarios. Although it might be not necessary to require the drivers to intervene on most occasions, it is still important to improve drivers' situation awareness (SA) in unexpected driving scenarios to improve their trust in and acceptance of AVs. In this study, we conceptualized SA at the levels of perception (SA L1), comprehension (SA L2), and projection (SA L3), and proposed an SA level-based explanation framework based on explainable AI. Then, we examined the effects of these explanations and their modalities on drivers' situational trust, cognitive workload, as well as explanation satisfaction. 
A three (SA levels: SA L1, SA L2 and SA L3) by two (explanation modalities: visual, visual + audio) between-subjects experiment was conducted with 340 participants recruited from Amazon Mechanical Turk. The results indicated that by designing the explanations using the proposed SA-based framework, participants could redirect their attention to the important objects in the traffic and understand their meaning for the AV system. This improved their SA and filled the gap of understanding the correspondence of AV’s behavior in the particular situations which also increased their situational trust in AV. The results showed that participants reported the highest trust with SA L2 explanations, although the mental workload was assessed higher in this level.
The results also provided insights into the relationship between the amount of information in explanations and modalities, showing that participants were more satisfied with visual-only explanations in the SA L1 and SA L2 conditions and were more satisfied with visual and auditory explanations in the SA L3 condition. Finally, we found that the cognitive workload was also higher in SA L2, possibly because the participants were actively interpreting the results, consistent with a higher level of situational trust. 
These findings demonstrated that properly designed explanations, based on our proposed SA-based framework, had significant implications for explaining AV behavior in conditional and highly automated driving.

\textbf{Keywords:} Explanations, Situation awareness, Modality, Automated driving.

%\vspace{20pt}
%\textbf{Pr\'ecis:} 

\newpage
\section{INTRODUCTION}
Automated vehicles (AV) have drawn broad interest. During the development of AV technology, artificial intelligence (AI) plays a fundamental role, but people still have difficulties in understanding or trusting the decisions made by AI due to its black-box nature \citep{Shen2020ToEO}. %In conditional and highly AVs, i.e., SAE (Society of Automotive Engineers) Levels 3 and 4 AVs, \citep{sae2021}, the drivers' responsibility as an active operator is switched to a passive passenger for the majority of the time, which reduces the SA of the driver and harms his/her performance when intervention is needed \citep{frison2019, endsley2019}. 
%According to Hancock's automation paradox, if systems are designed to rarely require people's intervention, they will rarely act (successfully) when it is required \citep{hancock2014}. This phenomenon might well describe the situation when people need to take over control of the AV when the AV cannot make safe decisions based on the perceived information. 
In conditional and highly AVs, i.e., SAE (Society of Automotive Engineers) Levels 3 and 4 AVs, \citep{sae2021}, the drivers' responsibility as an active operator is switched to a passive passenger for the majority of the time. This reduces driver’s SA since the attention mainly is switched to NDRT resulting in less eye-on-the-road time and harms his/her performance when intervention is needed\citep{frison2019, endsley2019}.  Clark et.al. \citeyearpar{ clark2017situational} showed that in unexpected takeover scenarios drivers who successfully took over the control within an acceptable time frame had a higher level of SA and responded faster than drivers who did not. 
%Furthermore, Merat et al. \citeyearpar{MERAT2014} found that drivers needed 35-40 seconds to take over control and stabilize the vehicle during SAE Level 3 vehicles and that drivers who were expecting the takeover requests performed better than those who did not pay attention to the driving situation. This may be explained by the driver's cognitive disengagement and passive information processing which makes difficult to maintain the sufficient level of SA while driving.  

When drivers are  out of the control loop, they will have a low level of SA, making it difficult for them to comprehend AV's behavior in unexpected situations. Moreover, it limits their ability to successfully take over control in critical situations, leading to accidents. For example, by analyzing Uber's AV fatal accident in Arizona \citep{uber2018}, it was revealed that the driver failed to take over control of the AV because she was engaged on her phone and was not aware of the pedestrian crossing the road. Regardless of who was responsible for the accident, such cases overall had negative impacts on trust in and public acceptance of AV. In particular, being unaware of the situation, drivers tend to interpret the AV's unexpected behavior as system malfunction that leads to trust issues in AVs. Hence, when the automated mode is on, the AVs should provide sufficient information to increase drivers' SA up to the “in-the-loop” level for proper understanding of the situation  and to ensure that the situation is under control. It is our belief, that improving the SA level will mitigate the unexpectedness and subsequent trust issues.

In complex intelligent systems, the lack of information about system behavior or misunderstanding of automation creates trust issues \citep{norman1990}, especially when the system acts outside of expectations. To foster trust in and acceptance of AV, it is crucial to make the system transparent for drivers and provide appropriate feedback on the system’s behavior. One of the concepts proposed to make black-box systems transparent is explainable artificial intelligence (XAI). It contributes to human-AI interaction by providing information about the main factors, which affect AI decisions and its future behavior. The AV, as a complex AI system, also needs to be explained for better human-AV team performance, since it is important to keep an appropriate level of trust in automation and effectively manage uncertainty. Previous studies already confirmed the necessity of feedback in autonomous driving \citep{Wintersberger2021, wiegand2020d, SEPPELT201966}. For example, \citet{Wintersberger2021} found that regardless of the trust in AV, people still preferred to be informed about forthcoming strategies and maneuvers.

Many human factors researchers made use of explanations of AVs' behavior and system feedback and status to help build the driver's mental model of the vehicle \citep{petersen2019situational,Koo2015WhyDM,koo2016understanding}. For example, \citet{Koo2015WhyDM} found that ``why'' (describing the reasoning for actions, e.g., ``obstacle ahead") information improved participants' understanding, trust, and performance, and ``why'' and ``how'' (describing actions, e.g., ``the car is breaking") information led to safest driving performance. Du et al. \citeyearpar{du2021designing} used explanations about future actions of the vehicle (i.e., ``what will'' information) and why the vehicle requested the driver to take over (i.e., ``why'' information) and the combination of the two during SAE Level 3 takeover transition periods. They found that ``what will'' information and ``what will'' + ``why'' information improved drivers' perceived ease of use and perceived usefulness, leading to potentially better takeover performance. These studies emphasized drivers' informational needs about the AV decisions and the driving scenarios during the takeover transition process. However, there is still no direct evidence to support that such information improved drivers' SA and eventually human-AV performance.

%In this study, we investigated how SA level-based explanations benefited human-AV interaction by using Endsley's SA model. Among numerous definitions of SA in the literature \citep{smithHancock1995, endsley1995, bendyMeister1999}, the most widely cited one was proposed by Endsley \citep{endsley1995}, which states that people process information in three hierarchical levels: 1) Level 1 SA: Perception of the elements in the environment, 2) Level 2 SA: Comprehension of the current situation, 3) Level 3 SA: Projection of future status in order to be up-to-date in the dynamic environment. Individuals need these three levels of SA in their decision-making process in complex dynamic human-machine interaction in various scenarios.
%Therefore, we hypothesize that explaining AV behaviors to accommodate drivers' informational needs in these three levels would result in different levels of understanding and human-AV performance. We designed a three by two between-subjects experiment, where three types of explanations were manipulated to provide three levels of SA with two modalities (visual, visual + auditory) across six scenarios. We examined the effects of explanations in the form of three levels of SA on drivers' situational trust, cognitive workload, and explanation satisfaction. 

\subsection{The present study}
As described above, previous studies addressed different issues in AVs (i.e., trust and takeover performance) through explanations, and provided important implications for designing AV systems. However, these solutions/models did not systematically assess how they improve drivers’ trust with a minimal level of cognitive workload. Therefore, it is necessary to frame the explanations theoretically to support human-AV interaction.

In this work, we proposed an SA-based explanation for the AV’s black-box system based on  Endsley \citeyearpar{endsley1995} and Sanneman and Shah \citeyearpar{sanneman2020SA}. First, we designed the explanations according to Endsley to support three levels of information process,  which states that people process information in three hierarchical levels: 1) Level 1 SA: Perception of the elements in the environment, 2) Level 2 SA: Comprehension of the current situation, 3) Level 3 SA: Projection of future status in order to be up-to-date in the dynamic environment. Individuals need three levels of SA in their decision-making process in complex dynamic human-machine interaction in various scenarios. Second, we designed the explanations to understand the decision-making process of the AV’s black-box system according to Sanneman and Shah's \citeyearpar{sanneman2020SA}’s mixed input/output principles   as follows: 1) ``what'' environmental input AV used to make a decision, 2) ``how'' the AV understands the input and ``how'' the input influences AV behavior and 3) ``what would happen'' if AV did not act in that way. 

We hypothesized that explaining AV behaviors to accommodate drivers' informational needs based on the above theories with  three levels of SA would result in different levels of understanding and human-AV performance. We expected that our explanation framework would foster trust with a relatively less increase in mental workload compared to the previous approaches due to the mapping of explanations to information processing levels. In order to test the hypothesis, we designed a three by two between-subjects experiment, where three types of explanations were manipulated to three levels of SA with two modalities (visual, visual + auditory) across six scenarios. We examined the effects of explanations in the form of three levels of SA on drivers' situational trust, cognitive workload, and explanation satisfaction.”

\section{Related Work}

\subsection{Explanations in AV}

In human factors research, explanations about the AV's behavior, system feedback and status, and driving scenarios were designed and provided to improve the transparency of system decisions and driver trust.  For instance, Wintersberger et al. \citeyearpar{wintersberger2019fostering} showed that augmented reality by coding traffic objects and future vehicle actions increased automation transparency and improved user trust and acceptance. \citet{Koo2015WhyDM} designed three different types of information to explain AV behavior about: 1) ``how'' the car was acting, 2) ``why'' the car was acting and 3) ``how'' + ``why'' the car was acting. Authors investigated AV-driver interaction in a scenario where the AV took control from the driver and suddenly braked to avoid collision with an obstacle. They explained the AV behavior before the AV started acting, and found that ``how'' + ``why'' information resulted in the safest AV-driver cooperation , but also produced the greatest cognitive workload than other explanations, which could lead to confusion and anxiety. The ``how’’  only information led to worse driving performance and unsafe cooperation since the drivers tried to take the control back from the AV but did not understand why the AV  behaved in that way.
Mackay et al.'s \citeyearpar{mackay2020} investigation into different amounts of feedback found that ``more information does not necessarily lead to more trust and may, in fact, negatively affect cognitive load''. Taehyun et al. \citeyearpar{taehyun2020} stated that type of explanation significantly affects trust in AVs and suggested an explanation format based on the attribution theory \citep{weiner1979theory}. They found that perceived risk moderated the effect of explanations on trust, i.e., attributional explanations led to the highest level of trust in low perceived risk compared to no or simple explanations.

In addition, the timing of the explanations (i.e., before or after particular action) also plays an important role in trust and acceptance in AVs. For example,  Körber et al. \citeyearpar{korber2018have} provided explanations of the causes of takeover requests after the takeover transitions, which led to no decrease in trust or acceptance, but improved participants' understanding of system behaviors.  \citet{Koo2015WhyDM} argued that explanations should be provided ahead of an event which also was supported by Haspiel et al.  \citeyearpar{haspiel2018explanations} and Du et. al. \citeyearpar{du2019look} studies, who found that explanations provided before the AV’s action promoted more trust than those provided afterward. Thus, it is recommended that we should provide explanations before the vehicle takes action. %and found that proposed a model of attributional explanations in AV and examined its effect on trust under different level of perceived risk. Results showed that attributional explanations had negative effect when the perceived risk was high compared to the other types of explanations. The level of trust on AV decreased as the perceived risk increased.

% ------ Modality researches ----
Other types of factors, such as forms, contents, and modalities of the explanations also play important roles in explanations in AVs. Wang et al. \citeyearpar{wang2020modality} explored how information modality influenced driver's performance and showed that both visual and auditory modalities had a significant influence, but on different aspects of driver's performance. In particular, visual information boosted performance efficiency and auditory information decreased reaction time. \citet{SEPPELT201966} showed that continuous feedback helped drivers to be involved in the loop of system performance and operations. Consistent with the multiple resource theory \citep{wickens2008multiple}, they found that the combined visual-auditory interface performed the best regarding drivers’ confidence and trust.

\subsection{Situation awareness and the out-of-the-loop problem}

\citet{merat2018OOTL} differentiated three kinds of loops in AV systems and described them as follows: 1) A driver was in the control loop when he/she was both in the physical control and monitoring the driving task, 2) a driver was on the control loop when the driver was only monitoring the driving task, and 3) a driver was out of the control loop as long as he/she was not monitoring the driving task. Thus, the out-of-the-loop problem in AVs describes the situation when the driver is not actively monitoring the system or the environment \citep{radlmayr2014}.  This issue is mostly due to driver's overtrust in AVs, since a certain level of “control” is needed to properly respond to situational changes or to reduce uncertainty in automated driving, such as monitoring and takeover control \citep{du2020psychophysiological,du2019examining,du2020examining}.

\citet{merat2018OOTL} emphasized that a key aspect to be in the control loop was the drivers' attention and cognitive responses to the changes in the system and in the dynamic environment, which was characterized by the driver's SA. In other words, when the driver is not in the control loop of the AV, the SA of system status and the driving environment may be reduced \citep{sebok2017implementing,Zhou:2019,zhou2021using}. Even if the driver is on the control loop (i.e., not in physical control of the vehicle, but monitoring the driving situation) \citep{merat2018OOTL}, he/she becomes a passive information processor, which would negatively affect the operator's understanding and comprehension (SA Level 2) of dynamic changes in the system even though the driver is aware of low-level information (SA Level 1) \citep{endsley1995outofl}. 
This is further aggravated by the black-box decision-making process of the AV and the monotonicity of automated driving, which lead to low vigilance and even drowsiness \citep{zhou2020driver,zhou2021predicting}.
However, SAE Levels 3-4 AVs allow drivers to conduct non-driving-related tasks without monitoring the driving task \citep{ayoub2019}. 
In order to resolve such conflicts (i.e., conducting NDRTs in AVs vs. requiring a certain level of SA in AVs), explanations are needed to help drivers resume their SA in time when a certain level of ``control'' or understanding is needed to respond the situational changes, especially during unexpected driving scenarios. %Thus, our study aimed to investigate the effects of different types of explanations in two modalities based on the concept of Endsley \citep{endsley1995}'s SA on their trust, cognitive workload, and satisfaction when the driver is out of the control loop but need to respond to different unexpected situational changes. 

%-----   \subsection{1.2. Situation Awareness-Based Framework for Explanation} -----
%The previous studies on explaining AV's behavior mainly focused on ``how'', ``why'' and  ``what'' information in order to make the system transparent. However, they represent already processed output from AI system while according to XAI principles before giving the system decision it is helpful to provide the inputs from environment that mostly contributed to the final decision, and the linkages between these objects that AI has created based on its knowledge. \citet{sanneman2020SA} proposed to combine XAI fundamentals and Endsley's SA levels \citep{endsley1995}, where the explanations leveraged the system inputs and outputs with three levels. They attempted to answer questions in three levels as follows: Level 1: ``What'' AI did or is doing; Level 2: ``Why'' the system behaves in that way or what is the significance of Level 1 information for the system; Level 3 ``what'' the system will do or what would happen ``if'' some of the inputs changed. However, the framework did not define which specific information should be given for XAI systems.

\section{METHOD}
% The data used in the development of algorithms was collected in two studies. The first study investigated the effects of cognitive load, traffic density, and takeover request lead time on takeover performance. The second study examined the effects of scenario type and vehicle speed on takeover performance. Also, participants in both experiments wore the same set of physiological sensors. 
% The similar experimental settings in both studies make it possible to combine the two datasets. At the same time, the varieties of takeover conditions from two studies increase model generalizability. 

\subsection{Participants}
In total, 340 participants (151 females and 189 males; Age =  39.0 $\pm$ 11.4 years old) in the United States participated in this study. All the participants were recruited from Amazon Mechanical Turk (MTurk) with a valid US driver's license. On average, participants had 15 ± 11.8 years of driving experience and the driving frequency was 5 ± 1 days per week. They were randomly assigned to one of the seven conditions as shown in Table \ref{table:conditions}, where L1, L2, and L3 conditions were mapped closely to three SA levels proposed by Endsley. More detailed information about the experiment conditions is described in the ``Scenario Design'' section. This study was approved by the Institutional Review Board at the University of Michigan.
%(i.e., Application number HUM00197521). 
Each participant was compensated with \$2 upon completion of the study. The average completion time of the survey was about 26 minutes across the conditions.

% \begin{table}[H]
% \caption{Experimental design with Modality and SA level as factors}
% \begin{center}
% %\vspace{-2mm}
% \captionsetup{justification=centering}
% \begin{tabular}{p{0.3\linewidth}p{0.25\linewidth}p{0.35\linewidth}} 
% \hline\hline
% Condition & \multicolumn{2}{c}{Independent variables}\\
% \hline
% Name & Modality & SA Level \\
% \hline
% Control: no explanation & - & -
% \\
% Text SA L1 & Visual & \multirow{2}{*} {Perception} \\ 
% Text + voice SA L1 & Visual + auditory 
% \\
% Text SA L2 & Visual & \multirow{2}{\linewidth}{Perception + Comprehension} \\ 
% Text + voice SA L2 & Visual + auditory 
% \\
% Text SA L3 & Visual & \multirow{2}{\linewidth}{Perception + Comprehension + Projection}
% \\ 
% Text + voice SA L3 & Visual + auditory \\
% \\
% \hline\hline
% \end{tabular}
% \end{center}
% \label{table:conditions}
% \end{table}

\begin{table}[H]
\caption{Experimental design with Modality and SA level as independent variables. The modality factor had two levels: 1) Visual, i.e., the explanation was given only in text format, and 2) Visual + Audio, i.e., the explanation was given in text and voice format simultaneously. The SA level factor had three levels: 1) SA L1, i.e., the explanation included only SA level 1 information (i.e., perception), 2) SA L2, i.e., the explanation included SA level 1 + level 2 information (i.e., perception and comprehension), and 3) SA L3, i.e., the explanation included SA level 1 + level 2 + level 3 information (i.e., perception, comprehension, and projection). Table cells represent the treated conditions in the experiment.}
% \begin{center}
\centering
%\vspace{-2mm}
\captionsetup{justification=centering}
\begin{tabular}{p{0.3\linewidth}p{0.25\linewidth}p{0.35\linewidth}} 
\hline\hline
\multirow{2}{\linewidth}{SA Level} & \multicolumn{2}{c}{Modality}\\ \cline{2-3}
& Visual & Visual + Audio \\
\hline
SA L1 & Text SA L1 & Text + audio SA L1 \\
SA L2 & Text SA L2 & Text + audio SA L2 \\
SA L3 & Text SA L3 & Text + audio SA L3 \\ 
% \hline
% \multicolumn{3}{c}{ \hfil Control: no explanation}\\
\hline\hline
\end{tabular}
\\[2ex]
% \end{center}
\begin{tablenotes}
   \item[*] A control condition was included in the experiment where participants did not receive any explanation.
  \end{tablenotes}
\label{table:conditions}
\end{table}

\subsection{Apparatus}
The study was conducted using a survey developed in Qualtrics (Provo, UT) and was published in MTurk. The survey was designed to evaluate the effects of SA and explanation modality on participants' situational trust, explanation satisfaction, and mental workload in uncertain situations while driving an AV. The driving scenarios were presented in videos created in the CarMaker autonomous driving simulation environment (Karlsruhe, DE).

\begin{table}[H]
\caption{Dependent variables}
\begin{center}
%\vspace{-2mm}
\captionsetup{justification=centering}
\begin{tabular}{p{0.30\linewidth}p{0.45\linewidth}p{0.25\linewidth}} 
\hline\hline
Measure & Description & Scale \\
\hline
% SA & (Manipulation check). Measured after watching first part of the simulation where explanations were provided. Participants answered all SA questions regardless of explanation level & SAGAT framework\\

Trust & Measured at the end of each scenario & STS-AD \\
Explanation Satisfaction & Measured at the end of each scenario & Explanation satisfaction scale \\

Mental Workload & Measured once participants watched all the 6 scenarios & DALI   \\

\hline\hline
\end{tabular}
\end{center}
\label{table:measures}
\end{table}

\subsection{Experimental design}
\textbf{Independent variables.} The experiment was a three (SA level: SA L1, SA L2, and SA L3) by two (modality: visual, visual + auditory) between-subjects factorial design with 6 scenarios. Alongside the 6 experimental conditions, a control condition with no explanations was also tested. The independent variables were the three levels of explanations mapped to three SA levels presented to the participants according to Endsley's SA model \citep{endsley1995} and in two types of modalities, i.e., visual and visual + auditory.
During the experiment, the participants' SA was measured through the Situation Awareness Global Assessment Technique (SAGAT) \citep{endsley1988design}. The SAGAT is a freeze-probe technique that requires pausing the simulation and asking a series of questions to assess the participants' awareness of the current situation. For each scenario, three different questions were developed to test the participants' perception of surrounding objects, comprehension of the current situation, and projection of the future state for that uncertain situation. All the questions designed for the SAGAT technique were developed based on a previous study \citep{VANDENBEUKEL2017302}. Table \ref{table:sagat} shows an example of multiple-choice questions for the training scenario (see Table \ref{table:scenarios}). Regardless of the experiment conditions, for each scenario, three SA questions were included in the survey corresponding to three levels of SA. The participants obtained one point if they answered the question correctly. With three questions for each scenario, the participants could get as many as 18 points, indicating perfect SA.

\begin{table}[H]
\caption{Example questions for the training scenario to measure SA with a SAGAT Questionnaire.}
\small
\centering
\begin{tabular}{p{2cm}p{5cm}p{7cm}c}
\hline\hline

Level of SA & Question & Options \\[0.5ex] 
\hline
Perception & 
The simulation just “froze”. Which road user was in front of the AV? 
&
  1) Bus, 2) \underline{Pedestrian}, 3) Cyclist, 4) I don't know, 5) Other
\\
 Compre-
 hension &  
 What caused you to seek your attention in this situation? & 
  1) \underline{Pedestrian’s intention to cross the street},
  2) Approaching heavy traffic,
  3) Approaching closed road,
  4) Faulty road lanes,
  5) I don't know,
  6) Other \\
 Projection &
 If the simulation resumes after this ``freeze'', what situation would require your extra attention or intervention? & 
  1) Other road user’s violations,  
  2) \underline{AV’s possibility to hit pedestrian},
  3) Impeding the traffic by stopping at intersection,
  4) I don't know,
  5) Other
 \\[1ex] 
 \hline\hline
\end{tabular}
\\[2ex]
\begin{tablenotes}
   \item[*] The underlined option indicates the correct answers.
  \end{tablenotes}
\label{table:sagat}
\end{table}

\subsection{Dependent measures}
The dependent variables in this study were situational trust, mental workload, and subjective satisfaction with explanations. Situational trust was measured by the self-reported Situational Trust Scale for Automated Driving (STS-AD) \citep{holthausen2020}. The model evaluates situational trust in six categories: trust, performance, non-driving related task (NDRT), risk, judgment, and reaction, by asking the following questions: 1) I trusted the automation in this situation, 2) I would have performed better than the AV in this situation, 3) In this situation, the AV performed well enough for me to engage in other activities, 4) The situation was risky, 5) The AV made a safe judgment in this situation, and 6) The AV reacted appropriately to the environment. All the six STS-AD scales were measured with a 7-point Likert scale. Situational trust was measured right after the participant watched one video that depicted a specific driving scenario. Thus, it was measured six times for six scenarios.

To understand the subjective satisfaction of the given explanations, the explanation satisfaction scale developed by Hoffman et al. \citeyearpar{hoffman2018} was used. In this study, it was presented to the participants with five items and was measured with a 7-point Likert scale. The following items were included: This explanation of how the AV behavior was 1) satisfying, 2) had sufficient details, 3) contained irrelevant details, 4) was helpful, 5) let me judge when I should trust and not trust the AV. Explanation satisfaction was also measured once right after the participant watched one specific driving scenario. Thus, it was measured six times for six scenarios.

The mental workload was measured using the driving activity load index (DALI) \citep{pauzie2008method}, which is a revised version of the NASA-TLX and specifically adapted to the driving tasks. DALI includes six factors: attention, visual, auditory, temporal, interference, and stress. In order to reduce the time of taking the survey, the cognitive workload was only measured once at the end of the survey using a 7-point Likert scale when the participants watched all the six scenarios. In the control and text-only scenarios, the auditory demand was removed.

\begin{figure}[bt!]
\centering
\includegraphics[width=1\linewidth]{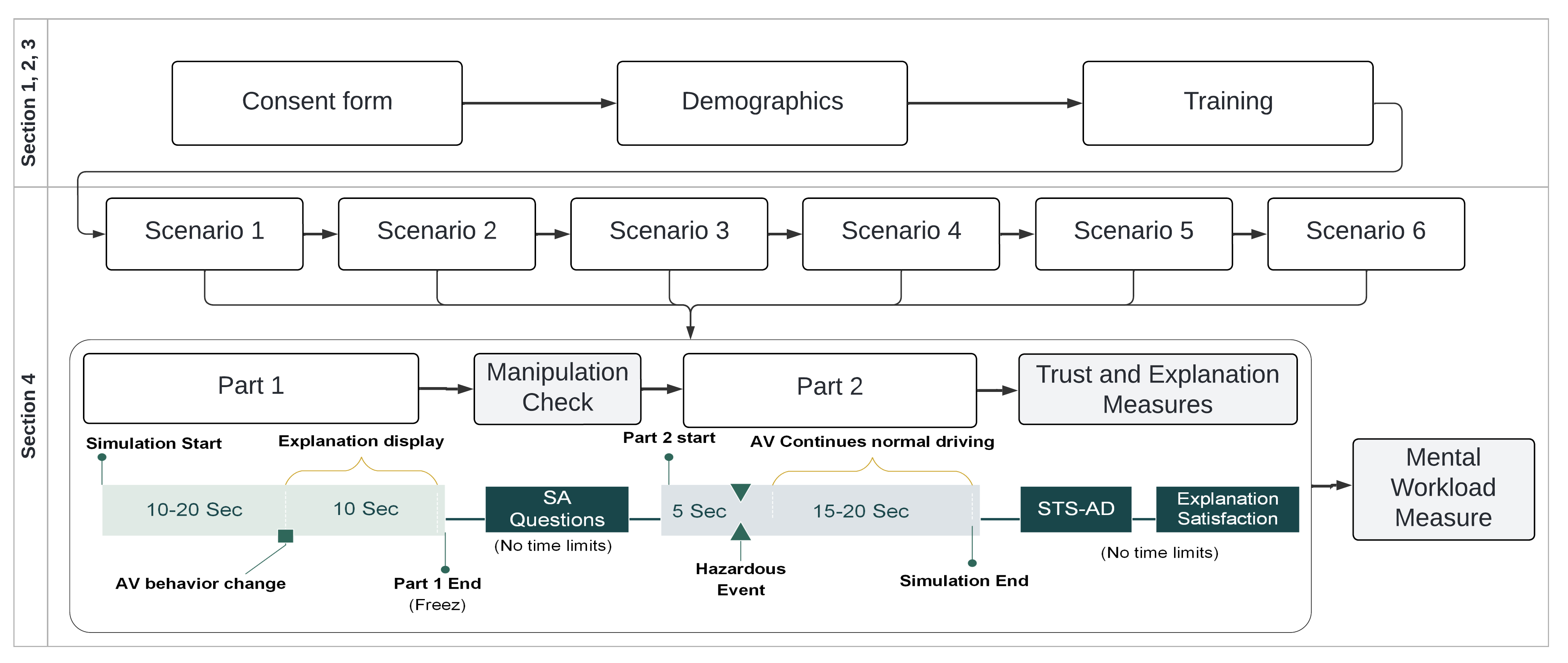}
\caption{Survey procedure.}
\label{fig:procedure}
\end{figure}

\begin{figure}%[ht!]
\centering
\includegraphics[height=3.5in, width=5.2in]{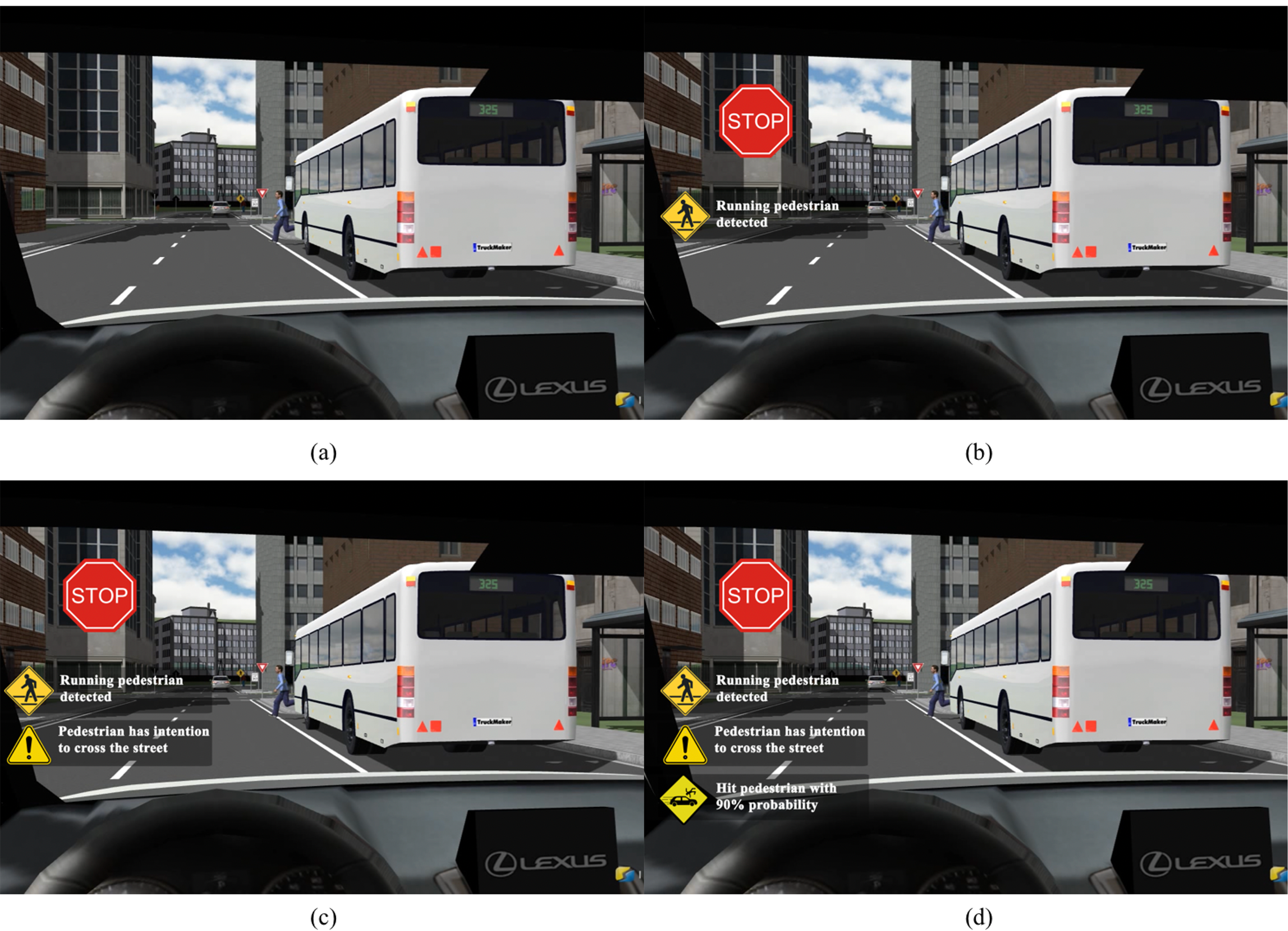}
\caption{Presented explanations S2 in (a) control, (b) SA L1, (c) SA L2 and (d) SA L3 conditions (see S3 L3: \url{https://youtu.be/GNL2cMK5Lyk}). }
\label{fig:visuals}
\end{figure}

\begin{table}[ht!]
\caption{Scenarios with description in this study}
\centering
\begin{tabular}{p{0.08\linewidth}p{0.18\linewidth}p{0.65\linewidth}}
 \hline
 \hline
Name & Scenario & Description and Link \\[0.5ex] 
 \hline
Training & Reluctant to turn right due to a pedestrian & City: The AV stops before turning right, and a pedestrian stands on the other side of the street and moves a little. There is no crosswalk. The AV slowly turns with intermittent stopping.
\url{https://youtu.be/B3Zw7-kZzoY} \\ 
S1 & Long wait at the intersection to turn
left & Highway: The AV approaches an intersection with a green traffic light. It stops behind the traffic light, and then moves a bit. After about 10 seconds, the AV finally turns left after an oncoming car passes. \url{ https://youtu.be/PfpsxPfmePg}\\
S2 & The AV stops and the pedestrian crosses & City: While driving, the AV stops abruptly. It waits. After seconds, a pedestrian crosses the street behind the bus. The AV continues driving. \url{https://youtu.be/i9nt3FvqbnM}\\
S3 & Unexpected stop due to an emergency vehicle & City: The AV stops. In some distance, there is a green traffic light. After a while, an emergency vehicle passes with the siren on. The AV waits for about 2 more seconds and continues driving. \url{ https://youtu.be/XmSrxEYeySo}\\
S4 & Strong and abrupt braking to reach the speed limit & City: The AV enters the city and brakes abruptly and strongly to reach the speed limit. \url{https://youtu.be/b5jrT4Mx9bg} \\
S5 & Early lane change due to heavy traffic & Highway: The AV changes to the right lane far away from the turn and it detects heavy traffic on the defined route. \url{ https://youtu.be/0kQw498WK20}\\
S6 & The AV waits for a long time before merging & Highway:  The AV slows down and stops. It needs to merge with the highway and waits for its chance with a safe distance while the AV's intention in merging lanes is not clear. Traffic is overloaded. \url{ https://youtu.be/L8I8ULMcuYw}
 \\[1ex] 
 \hline
 \hline
\end{tabular}
\label{table:scenarios}
\end{table}

\subsection{Survey Design and Procedure}
The survey consisted of four sections as illustrated in Figure \ref{fig:procedure}. The first section included a consent form. In the second section, the participants filled in a set of demographic questions. The third section was a training session, where the participants were given one simulation video example not used in the test session with three SA questions. Since the SA questions were designed based on the SAGAT technique, the freeze-probe technique was imitated for each scenario by dividing the simulation into two parts representing before and after the freeze situations. The fourth test section included six AV driving scenarios as shown in Table \ref{table:scenarios}. The participants watched the first part of each simulation video and answered three questions about their SA about the driving scenario (see Table \ref{table:sagat}). Then, they watched the second part of the video where they could see what happened actually. After each scenario, the participants evaluated their situational trust in AVs using the STS-AD scale and rated the given explanation(s) using the explanation satisfaction scale. After finishing all the six scenarios, the participants were required to report their mental workload about the explanations.

\subsection{Scenario Design}
Participants’ trust in AVs’ scenarios was investigated by manipulating their SA using three SA levels \citep{endsley1995} in different scenarios. All the situations were extracted from real driving scenarios and from Wiegand et al.'s work \citeyearpar{wiegand2020d}, where they explored the necessity of the explanations in unexpected situations while driving an AV. Seven scenarios were identified and simulation videos were created to visualize the situations (see Table \ref{table:scenarios}). In each scenario, the corresponding information was embedded into the video explaining the current situation before the AV started its actions. In this work, explanation modality was also explored by adding voice-over to simulations. In visual+auditory conditions, an auditory message with a synthesized female voice was added to provide the same situational explanations simultaneously with the visual explanations. Figure \ref{fig:visuals} illustrates the simulations for the S2 scenario (see Table \ref{table:scenarios}) correspondingly for the control, SA L1, SA L2, and SA L3 conditions. In the control condition, no explanation was given. The SA L1 condition provided information explaining the perception of the current environment, including the surrounding objects which influenced on the AV's behavior. In the SA L2 condition, additional information was used to explain how the AV understood the surrounding environment. The SA L3 condition included all the information from SA L2 and added extra information about how that might affect the AV's behavior in the future.

\subsection{Data Analysis}
Statistical analysis was conducted using the R language in RStudio. A two-way analysis of variance (ANOVA) was used to analyze the effects of the explanations on situational trust, explanation satisfaction, and mental workload. The alpha was set at 0.05 for all the statistical tests. Post-hoc analysis was conducted with Tukey's HSD test.

\section{RESULTS}
\subsection{Manipulation Check}
\begin{figure}
\centering
\includegraphics[width=0.6\linewidth]{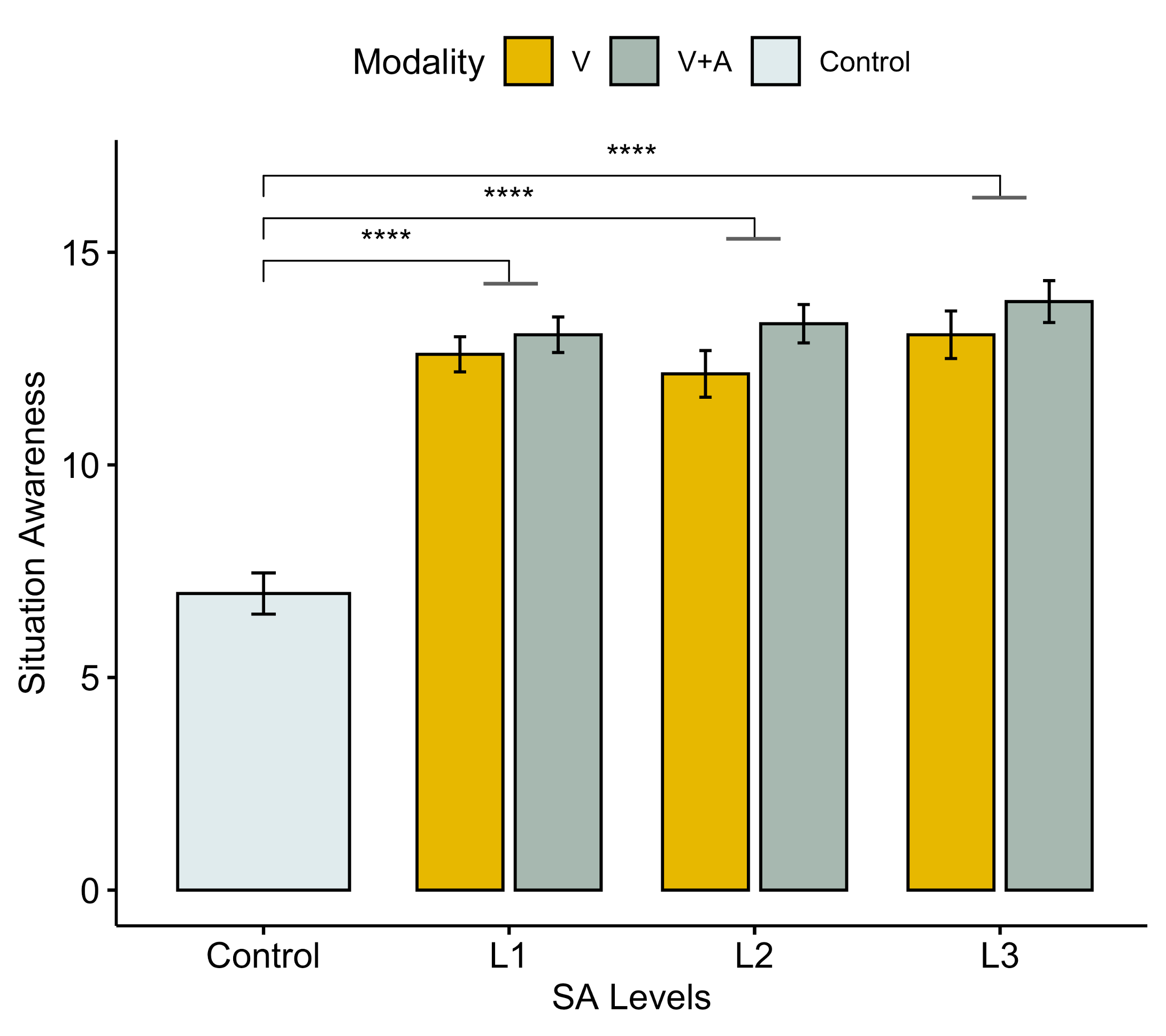}
\caption{Mean SA scores at different conditions and explanation modalities with standard error, where `***' indicates $p<0.001$.}
\label{fig:sa}
\end{figure}

In this study, the effect of the provided information on SA was explored with the control condition and three SA levels, where the participant’s SA was measured by the number of correct responses throughout the experiment. A two-way ANOVA test showed that there was a significant main effect of SA levels ($F(3,333)=38.23, p =.000, \eta^{2} = .253$) and modalities ($F(1,333)=4.26, p =.040, \eta^{2} = .009$) (see Figure \ref{fig:sa}). There was no significant interaction effect between SA levels and modalities ($F(2,333)= 0.28, p = .752$). The post-hoc analysis showed that SA was significantly higher in SA L1, L2, and L3 conditions compared to the control condition, and significantly higher in the visual + auditory modality ($p = .040$) compared to the visual-only modality. Figure \ref{fig:sa} illustrates the mean SA scores across different experimental conditions.

\subsection{Situational Trust}
\begin{figure}
\centering
\includegraphics[width=0.6\linewidth]{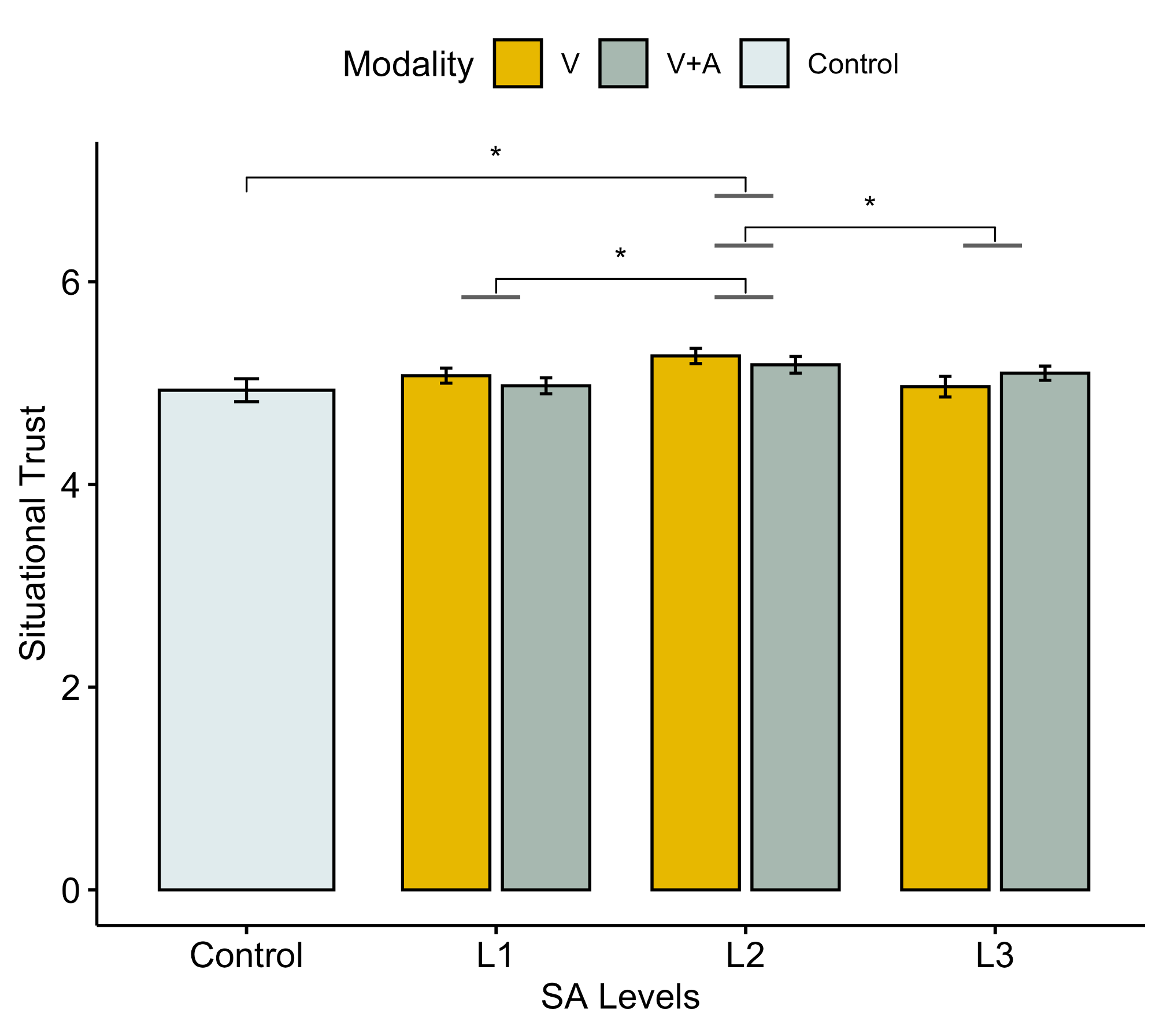}
\caption{Overall mean and standard error of situational trust measured by the SA levels and modalities, where `*' indicates $p<0.05$.}
\label{fig:trust}
\end{figure}

The means of the STS-AD over all six scenarios were calculated and analyzed with a two-way ANOVA. Results showed that the main effect of SA levels was significant ($F(2,294)=3.93, p =.020, \eta^{2} = .029$) whereas the main effect of modalities ($F(1,294)= .07, p = .789, \eta^{2} = .000$) and the interaction effect ($F(2,294)= 1.31, p = .272, \eta^{2} = .007$) were not significant (see Figure \ref{fig:trust}).
The post-hoc analysis showed that STS-AD in SA L2 was significantly higher than in SA L1 ($p = .036 $). Specifically, STS-AD in Text SA L2 was significantly ($p = .040 $) higher than that in Text + Voice SA L1.
And STS-AD was significantly higher ($p = .047 $) in SA L2 than that in SA L3.
Specifically, STS-AD in Text SA L2 was marginally ($p = .052$) higher than that in Text SA L3.
Compared to the control condition, it was found that only SA L2 was significantly higher ($p = .011$) mainly due to the visual-only modality ($p = .026$). As for the visual + auditory modality, the difference was not significant ($p = .131$).

\subsection{Explanation Satisfaction}
\begin{figure}[H]
\centering
\includegraphics[width=0.6\linewidth]{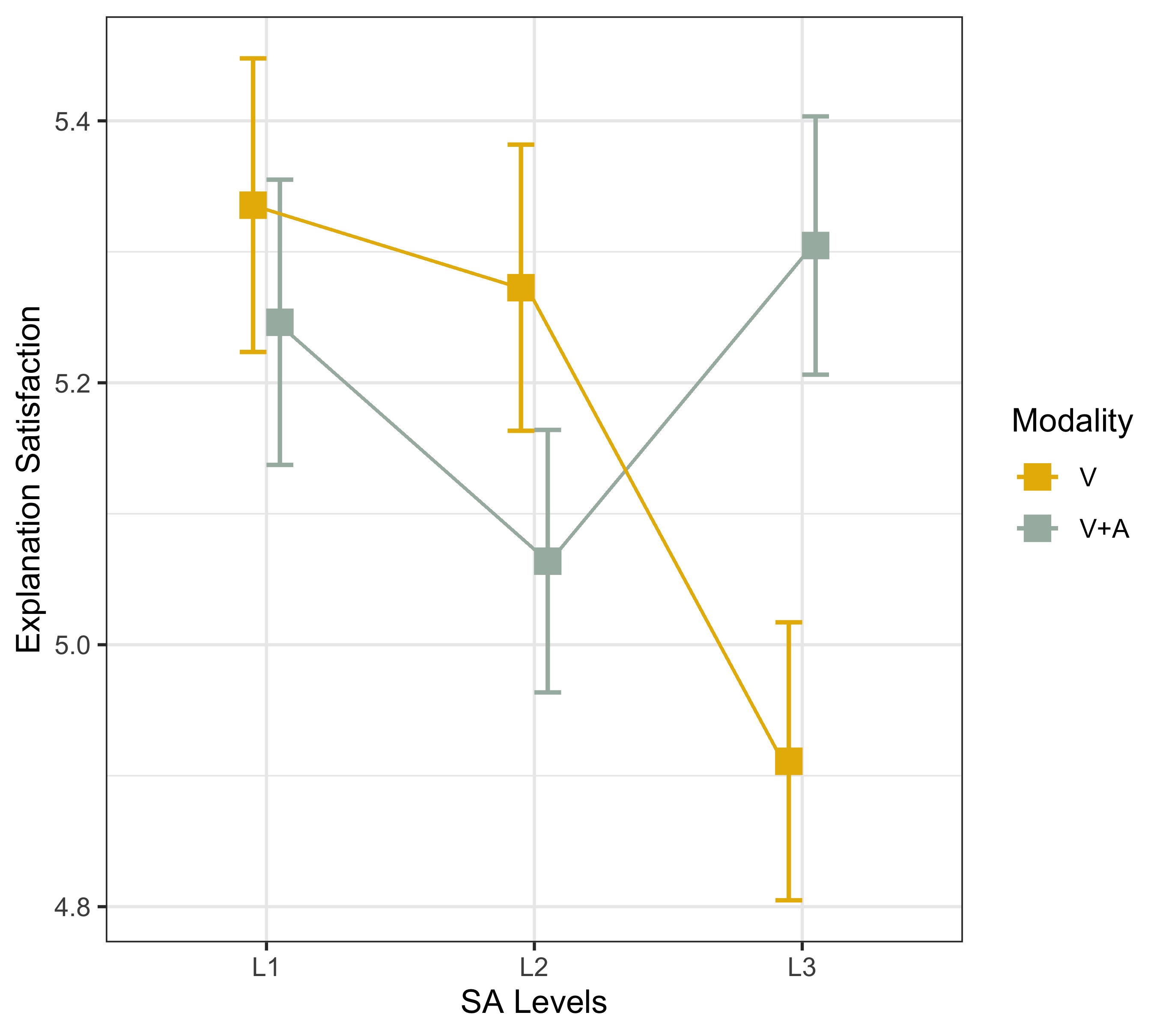}
\caption{Interaction effect of SA levels and modalities with standard error on explanation satisfaction.}
\label{fig:interaction}
\end{figure}

With regard to explanation satisfaction, the two-way ANOVA showed a significant interaction effect ($F(2,294)= 4.53, p= .012, \eta^{2} = .030 $). The post-hoc analysis showed that the participants were significantly more satisfied with the given explanations in the SA L1 ($p = .014 $) and SA L2 ($p = .043 $) conditions compared to the SA L3 condition when explanations were presented in the visual-only modality.
Furthermore, in the SA L3 condition, when a comparatively large amount of explanation information was presented, a significant effect of explanation modality was found that the visual + auditory condition resulted in a higher satisfaction score compared to the visual-only ($p = .009 $) condition (see Figure \ref{fig:interaction}).

\subsection{Mental Workload}

\begin{figure}[H]
\centering
\includegraphics[width=0.6\linewidth]{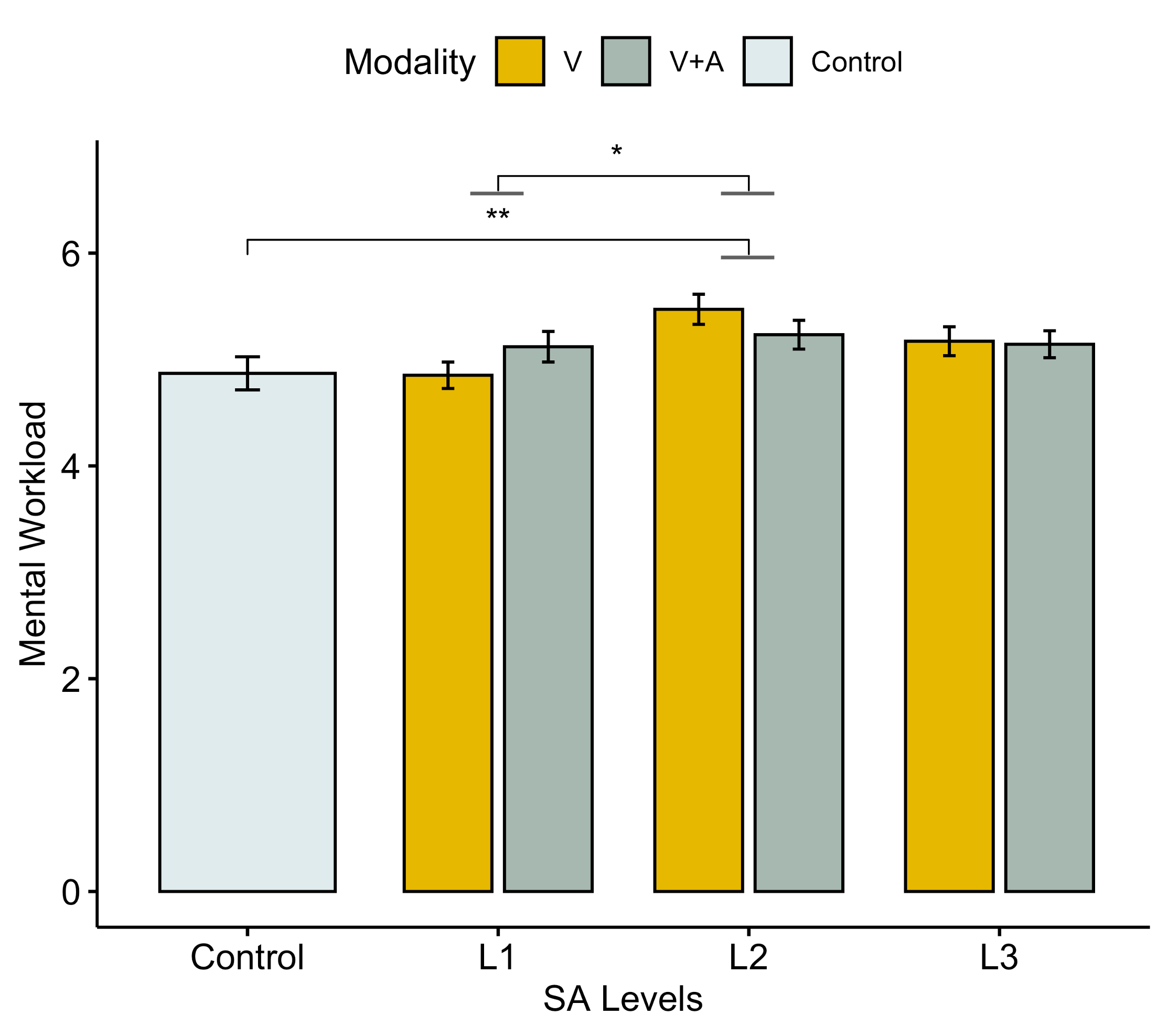}
\caption{Overall mean and standard error of mental workload measured by the SA level and modality, where `*' indicates $p<0.05$ and `**' indicates $p<0.01$.}
\label{fig:mw}
\end{figure}

The participants' self-reported mental workload was analyzed using the mean values of all the six DALI factors. %The effect of conditions were tested with two-way ANOVA
%with regression analysis ($R^2 = .024, F(3,296) = 2.457, p = .063$) 
As shown in Figure \ref{fig:mw}, we found a significant main effect of SA levels ($F(2,294)= 3.70, p= .026, \eta^{2} = .024 $) that participants' mental workload was significantly higher ($p = .018$) in the SA L2 condition than that in the SA L1 condition and than that in the control condition ($p = .009$). Specifically, we found that participants' mental workload in the Text SA L2 condition was significantly ($p = .016$) higher than that in the Text SA L1 condition and was significantly ($p = .012$) higher than that in the control condition. Thus, the significant differences were mainly caused by the visual-only modality.

\section{DISCUSSIONS}
\subsection{The Effects of SA}

% ------ Situational Trust --------------
In this study, we investigated the effects of SA explanations and modalities on situational trust, explanation satisfaction, and mental workload in AVs. First, our results partially supported that SA levels positively affected participants' situational trust  (see Figure \ref{fig:trust}) and SA L2 led to the highest level of situational trust. In this sense, situational trust appeared to be sensitive to SA. In particular, the participants' trust was significantly higher in SA L2 compared to SA L1 and L3, where the given information was either too little to foster the participants' perception and comprehension of the current situation or was redundant to notably improve trust  \citep{mackay2020}. 
One possible reason might be the out-of-the-loop problem, as Endsley et al. \citeyearpar{endsley1995outofl} found that SA L2 was the most negatively affected level by automation, where people's understanding of the situation significantly decreased, pushing them out of the control loop. When SA L2 explanations were provided to help the participants understand the situations and bring them back to the control loop, their situational trust was significantly improved.  Besides, consistent with Endsley \citeyearpar{endsley1995}, the participants might comprehend and project the future state at the same stage in SA L2, which indicates that the participants might already receive information that is supposed to receive in SA L3. For instance, in the scenario 2 (see Table \ref{table:scenarios}) comparing the SA L2 explanation (i.e., L1: “Running pedestrian detected”, L2: “Pedestrian has an intention to cross the street”), and SA L3 (i.e., L1, L2, and L3: “90\% risk of hitting a pedestrian”) explanations, the participants might project the risk of accident at L2, hence the L3 explanation was not useful. Therefore, there was also no significant difference between SA L2 and SA L3 in terms of cognitive processing as shown in Figure \ref{fig:mw}.

%In regard to modality, the participants' situational trust was higher when visual-only explanations were provided. This might be explained by the fact that the voice on the text explanation was generated by the machine (i.e., not natural enough) without any specific parameter configuration which might not match with in-car voice assistant system standards causing more distraction.% so that the participants could be easily got annoyed. 

%This is also helpful in explaining 
With regard to the interaction effect of SA levels and modalities on explanation satisfaction (see Figure \ref{fig:interaction}), the participants were more satisfied with the text explanations in SA L1 and L2 might be due to the machine-generated voice. As Tsimhoni, Green and Lai, \citeyearpar{tsimhoni2001listening} showed that natural speech led to a better comprehension of the given information compared to synthesized speech. However, participants were more satisfied with the combined visual and auditory explanations in SA L3.
This result was supported by the information processing theory \citep{wickens2008infprocessing} that it was easy to comprehend a large amount of information when more than one sensory resource (i.e., visual and auditory) was used while the participants might be annoyed to have redundant explanations with less information.

%\subsection{Mental Workload}
For cognitive workload, we found that participants had a higher cognitive workload in the SA L2 condition, especially the visual-only explanations, compared to the control and SA L1 conditions. One possible reason might be that the participants with explanations corresponding to SA L2 were actively interpreting the information to understand the driving scenarios, which improved their situational trust (see Figure \ref{fig:trust}). However, regardless of the extra information, SA L1 and SA L3 had similar levels of cognitive workload as the control group which might be due to the experiment design. %This might be explained by the experiment setting, i.e., the participants were not required to conduct any non-driving related tasks so that it was easy to consume the information for the participants in SA L1 while there was no need for the participants in SA L3 to estimate the future actions or to intervene appropriately in the driving scenarios.

\subsection{Implications}
We proposed to explain AV behavior based on the three levels of SA and XAI theoretically to satisfy their informational needs in unexpected scenarios, and empirically explored its effects on human-AV interaction. 
Considering the AV as a black-box AI system, the properly-designed explanations based on the SA framework  helped to define which components in the system should be explained to meet drivers' informational needs in order to understand AV's behavior. While previous studies have focused on ``how'', ``why'' and ``what'' information for explanations empirically \citep{Koo2015WhyDM,koo2016understanding,du2021designing}, this SA-based model focused more on XAI concepts and reduced the complexity of the situations to understand how the AI system came to that particular decision systematically.

During the interaction between the driver and the AV, it is important that the AV provides explanations with different levels of SA for the driver to understand its decision-making process. As pointed out by \citet{sanneman2020SA}, the key point is how to map such explanations into the needed three SA levels when designing such a black-box AV system as an XAI system. At SA level 1, we need to provide explanations about what objects are perceived from the environment to explain the effects of external factors on the decision-making process. At SA level 2, we should explain how the AV understands the situation by taking the perceived objects and their actions into consideration. At SA level 3, we might consider what actions would the AV and other road users take in the near future. Our explanations attempted to be designed based on the  theory-based SA model to satisfy drivers' informational needs and benefit them by improving their trust with a minimal level of cognitive workload.

%According to \citet{sanneman2020SA}, the proposed framework might help design the black-box AV system into an XAI system by mapping the explanations into three SA levels: 1) SA level 1, i.e., what objects are perceived from the environment to explain the effects of external factors on the decision-making process, 2) SA level 2, i.e., how the AV understand the situation by taking into the perceived objects and their actions into consideration based on the training patterns, and 3) SA level 3, i.e., what actions would the AV and other road users take in the near future. Our results showed that a theory-based SA model satisfied drivers' informational needs and benefited them by improving their trust with a minimal level of cognitive workload.

\subsection{Limitations and Future Work}
%talk about the major limitations of the study that can be used for future work
This study also has limitations that can be examined in future studies.
First, the experiment was conducted in a low-fidelity setting on MTurk due to the COVID-19 pandemic. The SA was measured with the SAGAT technique \citep{endsley1995} and we found that participants' SA was notably improved compared to the control condition. However, we could not identify significant differences among the three SA levels based on the provided explanations. One of the possible reasons might be that the data was collected on MTurk, where the scenarios were relatively short (30-45 seconds) and the fidelity was relatively low in the experiment. This potentially reduced the participants' engagement level. Another reason might be the absence of non-driving related tasks due to the difficulty in controlling participants when the experiment was conducted on MTurk, which allowed the participants to continuously monitor the ride. Nevertheless, the significant differences in SA between the control conditions and others indicated the importance of simple explanations in improving SA. Further investigations are needed to understand the effects of different explanations on SA and subsequently on trust, mental workload, explanation satisfaction, and the joint performance of the human-AV team in high-fidelity driving simulators.
Second, only self-reported measures were used to evaluate the trust and mental workload. Additional measures, such as physiological measures (e.g., galvanic skin response \citep{du2020psychophysiological}, eye-tracking \citep{de2019situation}) can be included in future studies.
Third, only a limited number of scenarios were tested in the experiment with low to moderate risks. Future studies can explore more scenarios with different levels of risk.
Fourth, since the experiment was conducted as a between-subjects design, the participants experienced only one of the SA levels, the results might be affected by individual differences and the low-fidelity of the experiment setting.

\section{CONCLUSION}

In this study, we designed an SA-based explanation framework to help drivers understand the driving situations and map the AV’s behavior properly to the situation. By exploring participants’ situational trust, cognitive workload, and explanation satisfaction, we evaluated the effectiveness of the framework in three SA levels and two modalities. Based on the results, it was partially supported that SA-based explanations improved participants’ situational trust. Among three levels, SA L2 resulted in higher situational trust and mental workload regardless of the explanation modality. However, modality preferences were changed from visual to visual and audio due to the explanation amount in SA L3. Overall, the results confirmed that the properly-designed explanations based on the SA-based framework helped orient drivers in the unexpected situation and assess the AVs’ behavior accurately leading to higher trust and acceptance of these vehicles.

\bibliography{HFES-bibliography}

% \newpage
% \section{Biographies}
% \textbf{Author 1} is a\\

% \textbf{Author 2} is a\\

% \textbf{Author 3} is a\\

\end{document}